\documentclass[%
superscriptaddress,
showpacs,preprintnumbers,
amsmath,amssymb,
 aps,
prl,
twocolumn,
10pt,
]{revtex4-1}

\usepackage{graphicx}
\usepackage{dcolumn}
\usepackage{bm}
\usepackage{hyperref}
\usepackage{textcomp}
\usepackage{siunitx}
\mathchardef\mhyphen="2D

\begin{document}


\title{Observation of the spin Peltier effect}


\author{J. Flipse}
\email[]{J.Flipse@rug.nl}
\author{F. K. Dejene}
\author{D. Wagenaar}
\affiliation{Physics of Nanodevices, Zernike Institute for Advanced Materials, University of Groningen, Nijenborgh 4,
9747 AG Groningen, The Netherlands.}
\author{G. E. W. Bauer}
\affiliation{Kavli Institute of NanoScience, Delft University of Technology, Delft, The Netherlands}
\affiliation{Institute for Materials Research and WPI-AIMR, Tohoku University, Sendai, Japan}
\author{J. Ben Youssef}
\affiliation{Universit\'{e} de Bretagne Occidentale, Laboratoire de Magn\'{e}tisme de Bretagne CNRS, 6 Avenue Le Gorgeu,
29285 Brest, France.}
\author{B. J. van Wees}
\affiliation{Physics of Nanodevices, Zernike Institute for Advanced Materials, University of Groningen, Nijenborgh 4,
9747 AG Groningen, The Netherlands.}


\date{\today}


\begin{abstract}
We report the observation of the spin Peltier effect (SPE) in the
ferrimagnetic insulator Yttrium Iron Garnet (YIG), i.e. a heat current
generated by a spin current flowing through a Platinum (Pt)$|$YIG
interface. The effect can be explained by the spin torque that transforms
the spin current in the Pt into a magnon current in the YIG. Via magnon-phonon
interactions the magnetic fluctuations modulate the phonon temperature
that is detected by a thermopile close to the interface. By finite-element
modelling we verify the reciprocity between the spin Peltier and spin
Seebeck effect. The observed strong coupling between thermal magnons and
phonons in YIG is attractive for nanoscale cooling techniques.
\end{abstract}

\maketitle

The discovery of the spin Seebeck effect (SSE) in YIG$|$Pt bilayers \cite%
{UchidaInsulator} opened up a new research direction in the field of spin
caloritronics. In the SSE a temperature difference between the magnons in
the magnetic insulator and the electrons in the metal contact leads to
thermal pumping of a spin current \cite{TheoryXiaoSSE, TheoryAdachiSSE,
TheoryTserkSSE}. In a suitable metal such as Pt this spin current is
transformed into an observable transverse voltage by the inverse spin Hall
effect \cite{SHE}. Numerical simulations of the phonon, magnon and electron
temperatures show good agreement with experiments \cite{SchreierSSE}. 
In this Letter we report the observation of the spin Peltier effect (SPE), 
which is the Onsager reciprocal \cite{Onsager} of the SSE.

The SPE is the generation of a magnon heat current
in the magnetic insulator by a spin current through the interface with the
metal contact. The latter can be generated by a charge current in the Pt
film that by the spin Hall effect generates a transverse spin current normal
to the interface. The spin Peltier heat current generates a temperature difference between magnons and
phonons in the YIG that when relaxing leads to a change in the lattice temperature. 
We confirm this scenario experimentally by picking up such temperature changes via proximity thermocouples. 
According to our modelling the experimental results
are consistent with Onsager reciprocity between the SPE and the SSE, which
we measure separately (see supplementary IV). Our results confirm recent indications for a strong magnon-phonon interaction in YIG at room
temperature \cite{TmpHillebrands, SchreierSSE, TmpRoschewsky}.

\begin{figure*}[tbp]
\includegraphics{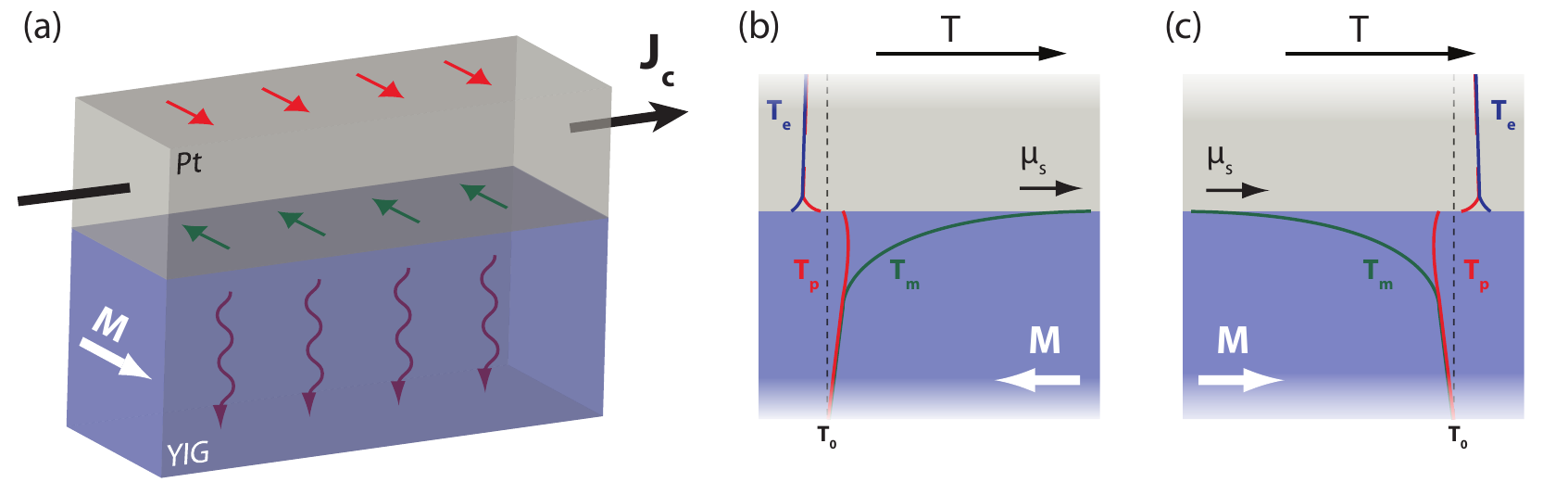}
\caption{(color online). Schematic figure of the spin Peltier effect at a Pt$%
|$YIG interface. (a) A charge current through the Pt creates a transverse
spin current induced by the spin Hall effect that generates a spin
accumulation $V_{s}$ at the boundaries. (b) When the spin magnetic moment \textit{$\mu_s$} is antiparallel to M
the spin torque transfers angular momentum and energy from the electrons in
the Pt to the magnons in the YIG thereby cooling the electrons and heating the magnons, effectively
raising the magnon temperature \textit{T$_{m}$} with respect to the electron
temperature \textit{T$_{e}$}. (c) When \textit{$\mu_s$} is parallel to M the spin torque
transfers angular momentum and energy from the magnons in the YIG to the
electrons in the Pt thereby cooling the magnons, effectively lowering \textit{T$_{m}$}
with respect to \textit{T$_{e}$}.}
\label{schematic_SP}
\end{figure*}

A charge current through a Pt strip generates a transverse spin current
induced by the spin Hall effect that leads to a spin accumulation \textit{$V_{s}$}
at the boundaries. At the interface to YIG the spin current is absorbed
as a spin transfer torque proportional to the spin mixing conductance \cite%
{MixC, Wang}, as depicted in Fig. \ref{schematic_SP}(a). When the magnetic moment of the spin accumulation (\textit{$\mu_s$}) 
at the Pt$|$YIG interface is parallel (antiparallel) to the
average magnetization direction, the spin torque transfers
magnetic momentum and energy from the electrons in the Pt to the magnons in
the YIG (or vice versa). Magnons are thereby annihilated (excited) (see Fig. \ref%
{schematic_SP}(b)) leading to cooling (heating) of the magnetic order
parameter (see Fig. \ref{schematic_SP}(c)). Since thermal magnons
equilibrate with the lattice by magnon-phonon scattering, the non-equilibrium magnons
affect the lattice temperature (see Fig. \ref%
{schematic_SP}(b) and (c)) depending on the
magnetization direction.

In the SSE \cite{TheoryXiaoSSE} the spin current density ($J_{s}$) pumped
from the YIG into the non magnetic metal is proportional to the temperature
difference between the magnons and electrons at the interface ($T_{m\text{-}%
e}=T_{m}-T_{e}$) and the interface spin Seebeck coefficient $L_{S}$, $%
J_{s}=L_{S}T_{m\text{-}e}$. In order to arrive at a symmetric linear
response matrix that reflects Onsager symmetry, the sum of the products of
currents and driving forces should be proportional to the dissipation \cite%
{OnsagerR}, leading to (see supplementary I) 
\begin{equation}
\left( 
\begin{array}{c}
J_{s} \\ 
Q_{m\text{-}e}%
\end{array}%
\right) =\left( 
\begin{array}{cc}
g_{S} & L_{S}T \\ 
L_{S}T & \varkappa _{S}^{I}T%
\end{array}%
\right) \left( 
\begin{array}{c}
V_{s}/2 \\ 
T_{m\text{-}e}/T%
\end{array}%
\right)   \label{eq:model}
\end{equation}%
Here we used the Onsager Kelvin relation $\Pi
_{S}=S_{S}T=L_{S}/ \left( g_{S}T\right),$ where the spin Seebeck $S_{S}=\left(
dV_{s}/2dT\right) _{J_{s}=0}$ and spin Peltier $\Pi _{S}=\left( dQ_{m\text{-}%
e}/dJ_{s}\right) _{\partial T_{m\text{-}e}=0}$ coefficients have been defined. $g_{S}$ is
the average spin conductance per unit area when spin accumulation and magnetization are
collinear, i.e. the \textit{$V_{s}$} at the YIG$|$Pt interface is either parallel or
antiparallel to the average YIG magnetization. $g_{S}\approx 0.16g_{r}$ at room temperature 
\cite{G_S}, where $g_{r}$ is the real part of the spin-mixing
conductance per unit area. $\varkappa _{S}^{I}$ is the magnetic contribution to the
interface heat conductance per unit area \cite{SchreierSSE}. The SPE heat current density we set
out to discover is therefore 
\begin{equation}
Q_{m-e}=L_{S}T\frac{V_{s}}{2}.  \label{eq:SPE}
\end{equation}

\begin{figure}[tbp]
\includegraphics{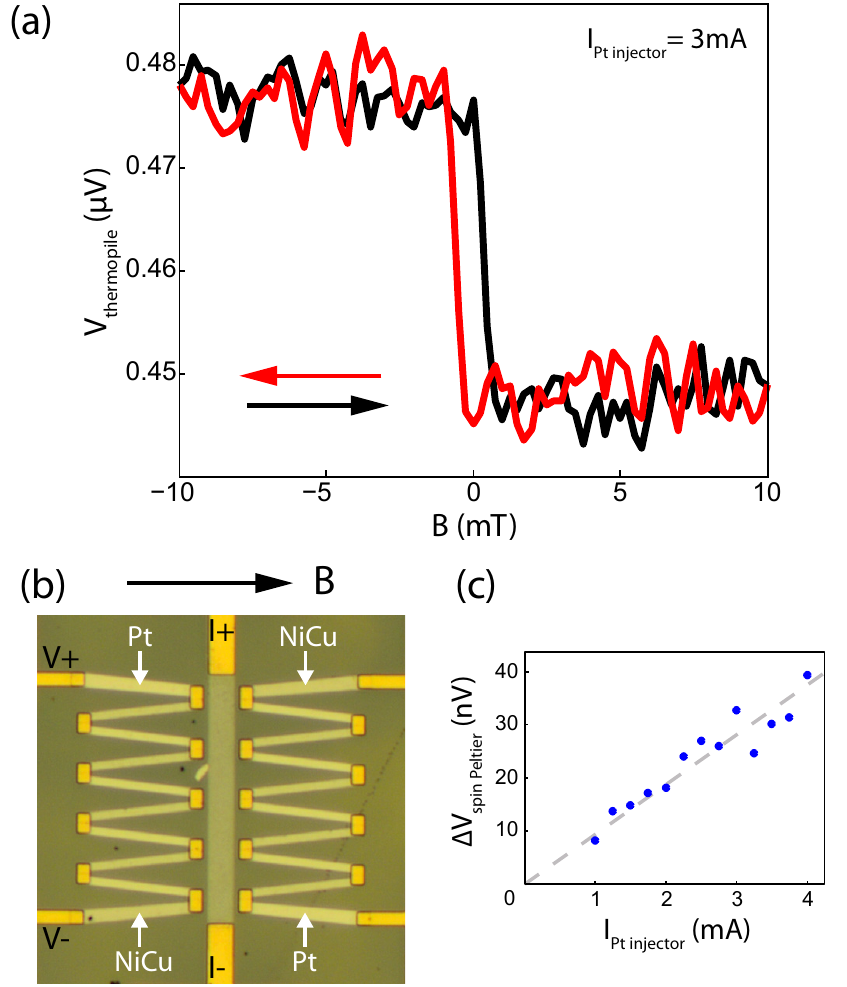}
\caption{(color online). (a) First harmonic voltage across the thermopile as
a function of applied magnetic field. The difference between the voltage at
positive and negative fields is the spin Peltier signal. (b) Optical
microscope picture of the device. (c) The spin Peltier signal ($%
\Delta V_{\text{spin Peltier}}$) as a function of the charge current through
the Pt injector. }
\label{data_plots}
\end{figure}

The devices designed for observing the SPE are fabricated on top of a 200 nm
thick single-crystal (111) Y$_{3}$Fe$_{5}$O$_{12}$ (YIG) film grown on a \SI{500}{\micro\m} 
thick (111) Gd$_{3}$Ga$_{5}$O$_{12}$ (GGG) substrate by
liquid-phase-epitaxy. Two temperature sensors are fabricated in close
proximity to the Pt$|$YIG interface. The optical microscope image in Fig. %
\ref{data_plots}(b) shows the 20$\times $\SI{200}{\micro\m}$^{2}$ and 5
nm thick Pt injector film. The thermopile sensors consist of five 40 nm
thick Pt-Constantan (Ni$_{45}$Cu$_{55}$) thermocouples in series that are
very sensitive because of the large difference in the Seebeck coefficient
of these metals. In the thermopile on the right of the Pt injector
the Pt$|$Ni$_{45}$Cu$_{55}$ order is reversed for additional cross
check measurements. The two thermopiles and the Pt injector are connected
to 5$|$100 nm thick titanium$|$gold contacts, providing good
thermal anchoring and electrical contact to bonding pads \SI{30}{\micro\m}
away. All structures are patterned by electron beam lithography. The Pt
injector and the Ni$_{45}$Cu$_{55}$ are deposited by DC
sputtering while electron beam evaporation has been used to make the Au
contacts and Pt thermocouple components.

An AC current is sent through the Pt injector, from I$^{+}$ to I$^{-}$ (Fig. %
\ref{data_plots}(b)), to create $V_{s}$. The voltage over the thermopile (V$%
^{+}$ and V$^{-}$) is simultaneously recorded. Using a standard lock-in
detection technique the first harmonic response ($V\propto I$) is extracted
from the measured voltage. A low excitation frequency of 17 Hz was used to
ensure a thermal steady-state condition. All measurements are carried out at
room temperature.

In Fig. \ref{data_plots}(a) the first harmonic voltage over the thermopile
is shown as a function of an applied in-plane magnetic field (\textit{$B$}) for a
root-mean-square current of 3 mA through the Pt injector. A clear switch is
observed just after the applied field becomes positive, in line with the
magnetization reversal of YIG at very small coercive fields \cite{VSM}. The signal switches back to its original value when reversing the field with a small hysteresis. We
measure a SPE signal of 33 nV on top of a background voltage of \SI{0.463}{\micro\V}. 
We observe linear scaling of the SPE signal for currents between 1 and 4 mA in the Pt (\textit{I$_{\textrm{Pt injector}}$})(see Fig. \ref{data_plots}%
(c)). Results for four different samples (from two
different batches) match the signal presented here within 15 \%. The
measurements were repeated with \textit{B} rotated 90$^\circ$. No SPE signal was
observed in this configuration while the background remained the same (see
supplementary II), which confirms our interpretation.

In order to obtain quantitative information we carry out 3D finite element modelling of our devices \cite%
{comsol}. As discussed above, the SPE heat current ($Q_{m\text{-}e}$) flows between the electron and magnon systems through the Pt$|$YIG
interface. $Q_{m\text{-}e}$ is calculated using Eq. (\ref{eq:SPE}) and 
\begin{equation}
V_{s}=\theta J_{c}\cdot \eta \cdot \tanh \left( \frac{t}{2\lambda }\right) 
\label{eq:Vs}
\end{equation}%
where $\theta $ is the spin Hall angle, $t$ the Pt film
thickness, $J_{c}$ the charge current density through the Pt injector, $\rho 
$ the Pt resistivity, $\lambda $ the spin-flip diffusion length and $\eta
=2\lambda \rho \cdot \lbrack 1+g_{S}\rho \lambda \coth \left( \frac{t}{%
\lambda }\right) ]^{-1}$ \vspace{0.1cm}a backflow correction factor. The
heat charge current densities in Pt are modelled by a three reservoir
model of thermalized phonons, magnons and electrons at temperatures \textit{$T_{ph}$}, \textit{$T_{m}$} 
and \textit{$T_{e}$}, respectively \cite{SchreierSSE}. In linear response the
charge ($J_{c}$) and heat ($Q$) current densities in the bulk of the
materials are related to their driving forces, i.e. gradients
of ($V$, $T_{ph}$, $T_{m}$ and $T_{e}$) as $\vec{Q}_{x}=\kappa _{x}\vec{%
\nabla}T_{x}$ and%
\begin{equation}
\left( 
\begin{array}{c}
\vec{J}_{c} \\ 
\vec{Q}_{e}%
\end{array}%
\right) =-\left( 
\begin{array}{cc}
\sigma  & \sigma S \\ 
\sigma ST & \kappa_{e}
\end{array}%
\right) \left( 
\begin{array}{c}
\vec{\nabla}V \\ 
\vec{\nabla}T_{e}
\end{array}%
\right) 
\end{equation}%
where $x$ is $ph$ or $m$, $\sigma $ is the electrical conductivity, $S$ the Seebeck coefficient
and $\kappa _{ph}$, $\kappa _{m}$ and $\kappa _{e}$ are the phonon, magnon
and electron thermal conductivities, respectively. The interaction between
the magnon and phonon subsystems in YIG and between the phonon and
electron subsystems in Pt are taken into account by using
thermal relaxation lengths, $\lambda _{m-ph}$ and $\lambda _{e-ph},$
respectively (see supplementary III), 
\begin{equation}
\nabla ^{2}T_{m\text{-}ph}=\frac{T_{m\text{-}ph}}{\lambda _{m\text{-}ph}^{2}}%
\qquad \text{and}\qquad \nabla ^{2}T_{e\text{-}ph}=\frac{T_{e\text{-}ph}}{%
\lambda _{e\text{-}ph}^{2}}.  \label{eq:Tmprelaxation}
\end{equation}%
The phonon interface heat conductance ($\kappa _{ph}^{I}$) and heat exchange
between magnons and electrons across the interface ($\varkappa _{S}^{I}$) are
treated as boundary conditions \cite{SchreierSSE} (see Supplementary III).

\begin{table}[tbp]
\caption{Material parameters used in the model. Both $\protect\sigma$ and S
are measured in separate devices \protect\cite{SandSigma} except for $%
\protect\sigma$ of the Pt injector, which is extracted from the SPE devices
directly. $\frac{\protect\kappa_{ph}}{\protect\kappa_{e}}$is adopted from
Ref. \onlinecite{SchreierSSE} and the total $\protect\kappa = \protect\kappa%
_{ph} + \protect\kappa_{e}$ is calculated using $\protect\kappa = \frac{%
\protect\sigma}{\protect\sigma_{bulk}}\protect\kappa_{bulk}$.}
\label{table:model}
\begin{ruledtabular}
\begin{tabular}{c c c c c}
\renewcommand{\arraystretch}{2}
 & $\sigma$ & S & $\kappa_{ph}$ & $\kappa_{e}$\\
 & (S/m) & (\SI{}{\micro\V}/K) & (W/(m$\cdot$ K)) & (W/(m$\cdot$ K))\\
\hline
YIG & - & - & 6 & - \\
GGG & - & - & 8 & - \\
Au & 2.7$\cdot$10$^7$ & 1.7 & 1 & 179 \\
Pt injector & 3.5$\cdot$10$^6$ & -5 & 3 & 23 \\
Pt thermocouple & 4.2$\cdot$10$^6$ & -5 & 4 & 28 \\
Ni$_{45}$Cu$_{55}$ & 1$\cdot$10$^6$ & -30 & 1 & 9\\
\end{tabular}
\end{ruledtabular}
\end{table}

\begin{figure*}[tbp]
\includegraphics{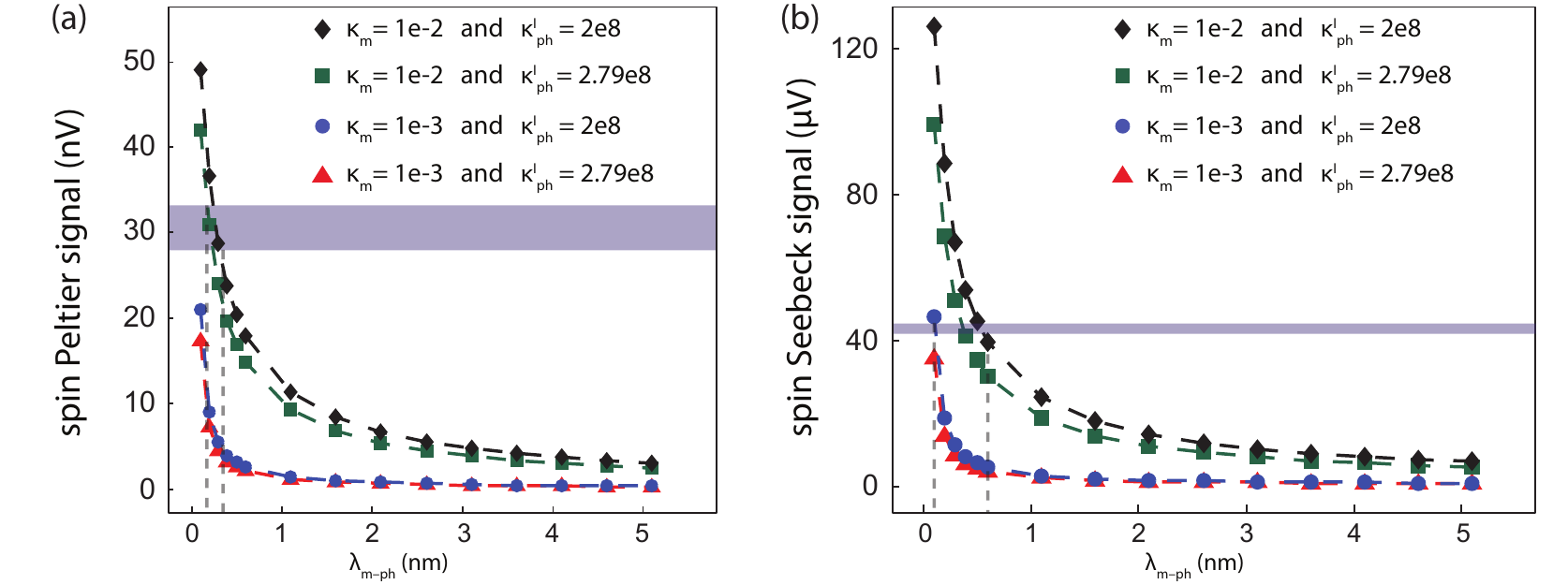}
\caption{(color online). The modeled SPE (a) and SSE (b) signal versus $%
\protect\lambda_{m\mhyphen ph}$ for two different values of $\protect\kappa%
_{m}$ (W/(m K)) and two different values of $\protect\kappa^I_{ph}$ (W/(m$^2$K)). 
The semi transparent blue bar indicates the range of measured SPE and
SSE effect signals. }
\label{fig:plots_model}
\end{figure*}

This model is evaluated for the material parameters listed in table \ref%
{table:model}. Additionally, we adopt $g_{r}=7\times 10^{14}$ $\Omega ^{-1}$m$%
^{-2}$ \cite{MixCMeas}, $\theta =0.11$ 
\cite{SchreierSSE}, $\lambda =1.5$ nm \cite{SchreierSSE} and $L_{S}=7.24\times 10^{9}$ A/(m$^2$K) \cite%
{TheoryXiaoSSE, SchreierSSE}. The magnon heat conductivity of YIG ($\kappa _{m}$) at room
temperature is not well known so we used a $\kappa _{m}$ of 10$^{-2}$ and 10$%
^{-3}$ W/(mK) in order to cover the range of estimated values \cite%
{SchreierSSE, KMagnon}. For Pt$|$YIG a $\kappa _{ph}^{I}$ of $%
2.78\times 10^{8}$ W/(m$^{2}$K) obtained from the acoustic mismatch model was adopted
\cite{SchreierSSE}. Since this model tends to overestimate the heat
conductance \cite{IRes}, we also used $2\times 10^{8}$
W/(m$^{2}$K). In figure \ref{fig:plots_model}(a) the results are shown as a function of 
$\lambda _{m\text{-}ph}$. The semi transparent blue horizontal bar indicates the range
of measured SPE signals that are best fitted by a $\lambda _{m\text{-}ph}$
of 0.1 to 0.2 nm for the ranges of $\kappa _{m}$ and $\kappa _{ph}^{I}$
discussed above.

SSE samples were fabricated and simulated by the same model and parameters
used above (see Supplementary IV). In Fig. \ref{fig:plots_model}(b) the
results are plotted and best fitted by $\lambda _{m\text{-}ph}$ between 0.2 and 0.5 nm, 
which is consistent with the values found for the SPE, as is
indeed required by Onsager reciprocity. This implies that our model captures
the essential physics of the interacting electron, magnon and phonon
systems.

The observed SPE signal in Fig. \ref{data_plots}(a) corresponds to a phonon temperature difference of 0.25 mK at the thermopile, 
which according to the model is 39 \% of the phonon temperature difference directly at the Pt$|$YIG
interface. By engineering devices in which the phonon heat loss through the
substrate is minimized by thinner or etched YIG films could
therefore significantly enhance the measured signal. Altering the Pt
injector coupling to the heat sink or placing the thermocouple on top of the
Pt injector might also help.

The $\lambda _{m\text{-}ph}$ found here is smaller than the one adopted by
Ref. \onlinecite{SchreierSSE} ($\approx $ 6nm) by roughly an order of
magnitude. Actually Schreier \textit{et al}.'s simulations might agree
better with their measurements for smaller values as well. $\lambda _{m\text{-}ph}$ extracted from Fig. \ref{fig:plots_model} is quite
sensitive to small variations in the modelling, which implies a large
uncertainty. Nevertheless even when accepting a large error bar from 0.1 to
6 nm for $\lambda _{m-ph}$ we may conclude that thermal magnons and phonons interact strongly \cite{TmpHillebrands}.

The background signal in the SPE data is a factor 20 higher than we would
expect from conventional charge Peltier heating and cooling at the Au$|$Pt
injector interfaces. Reference measurements on the second thermopile on the
other side of the Pt injector excludes charge current leakage to the
thermopile. For an identical configuration, V$^{+}$ on the same side as I$%
^{+}$, we find an opposite sign of the measured voltage, as expected for a thermal signal since the Pt$|$Ni$_{45}$Cu$_{55}$
thermopile sequence is inverted. A current leak would not change sign and
can therefore be excluded. The background in the second harmonic voltage is
likely to be caused by the thermovoltage across the thermopile due to Joule
heating in the Pt injector, since its value agrees within 17\% with the
modeled one. Additional measurements of frequency dependent properties (see
Supplementary V) rule out  pick-ups due to capacitive or inductive
couplings.

We checked that the sign of the experimentally observed SPE and SSE signals obey reciprocity. Furthermore the voltage measured across the Pt detector
in a RF spin pumping measurement matches the sign of the SSE voltage for the same 
geometry when heating the YIG relative to the Pt, as previously reported \cite{SSE_SP_Hill, SSE_L_YIG_2}. However, the absolute sign of these three effects is still
under investigation.

In conclusion, we report experimental proof that a spin accumulation at a Pt$%
|$YIG interface induces heat exchange between  electrons and magnons on both
sides. Using thermal modelling to knit the theory of interface transport to
the observables we demonstrate that the SPE is the Onsager reciprocal of the
SSE and confirm a strong interaction between thermal magnons and phonons in
YIG, as reported earlier \cite{TmpHillebrands}. We hope that these results
can contribute to a better understanding of coupling between thermomagnetic
and thermoelectric properties. Our proof of principle opens new strategies
for nanoscale cooling applications.

We would like to acknowledge B. Wolfs, M. de Roosz and J. G. Holstein for
technical assistance. This work is part of the research program of the
Foundation for Fundamental Research on Matter (FOM) and supported by NanoLab
NL, Marie Curie ITN Spinicur, DFG Priority Programme 1538 "Spin-Caloric Transport", Grant-in-Aid for
Scientific Research A (Kakenhi) 25247056 and the Zernike Institute for
Advanced Materials.

\onecolumngrid
\clearpage

\section{SUPPLEMENTARY INFORMATION}
\section{I. O\lowercase{nsager reciprocity for the spin }S\lowercase{eebeck and spin }P\lowercase{eltier effect}}

The linear response matrix of thermoelectrics reflects Onsager reciprocity
when the sum of the products of currents times driving forces equals the
dissipation \cite{OnsagerR}. When $I_{c}$ and $Q$ are the charge and heat
currents driven by voltage and temperature differences $\Delta V$ and $%
\Delta T$, $\dot{F}=I_{c}\Delta V+Q\Delta T/T$ equals the dissipation and we
obtain the symmetric response matrix \cite{OnsagerR}: 
\begin{equation}
\left( 
\begin{array}{c}
I_{c} \\ 
Q%
\end{array}%
\right) =-\left( 
\begin{array}{cc}
G & GST \\ 
GST & KT%
\end{array}%
\right) \left( 
\begin{array}{c}
\Delta V \\ 
\Delta T/T%
\end{array}%
\right)   \label{eq:Onsager_Conv}
\end{equation}%
where $G$ is the electrical conductance, $S$ the Seebeck coefficient, and $K$
the heat conductance. Here the Onsager-Thomson relation for the Peltier
coefficient $\Pi =ST$ has already been implemented.

In the case of the spin Seebeck effect (SSE) and the spin Peltier effect
(SPE) for magnetic insulators, the spin accumulation at the interface drives
a spin current. To ensure reciprocity, we have to compute the Joule heating
caused by the spin currents: 
\begin{equation}
\dot{F}=I_{\uparrow }\Delta V_{\uparrow }+I_{\downarrow }\Delta
V_{\downarrow }=G_{\uparrow }\Delta V_{\uparrow }^{2}+G_{\downarrow }\Delta
V_{\downarrow }^{2}  \label{spin_dissipation}
\end{equation}%
where the subscripts denote the up and down spin contribution. Comparing
this with the product of $I_{s}$ with $\Delta V_{s}$: 
\begin{equation}
I_{s}\Delta V_{s}=(G_{\uparrow }\Delta V_{\uparrow }-G_{\downarrow }\Delta
V_{\downarrow })(\Delta V_{\uparrow }-\Delta V_{\downarrow })=2(G_{\uparrow
}\Delta V_{\uparrow }^{2}+G_{\downarrow }\Delta V_{\downarrow }^{2})
\label{spin_dissipation2}
\end{equation}%
we conclude that $\Delta V_{s}/2$ is the proper driving force. The linear
response relations for spin and heat current \textit{densities} at the interface between a normal metal and a ferromagnet
then reads (Eq. (1) of the main text, where all variables and parameters are
introduced):%
\begin{equation}
\left( 
\begin{array}{c}
J_{s} \\ 
Q_{m-e}%
\end{array}%
\right) =\left( 
\begin{array}{cc}
g_{S} & L_{S}T \\ 
L_{S}T & \varkappa _{S}^{I}T%
\end{array}%
\right) \left( 
\begin{array}{c}
V_{s}/2 \\ 
T_{m\text{-}e}/T%
\end{array}%
\right)  \label{eq:Onsager_SPE}
\end{equation}
and we omitted the charge sector because we are dealing with a ferromagnetic insulator.

\section{II. M\lowercase{easurement with} B \lowercase{rotated 90$^\circ$}}

\begin{figure}[tbp]
\includegraphics{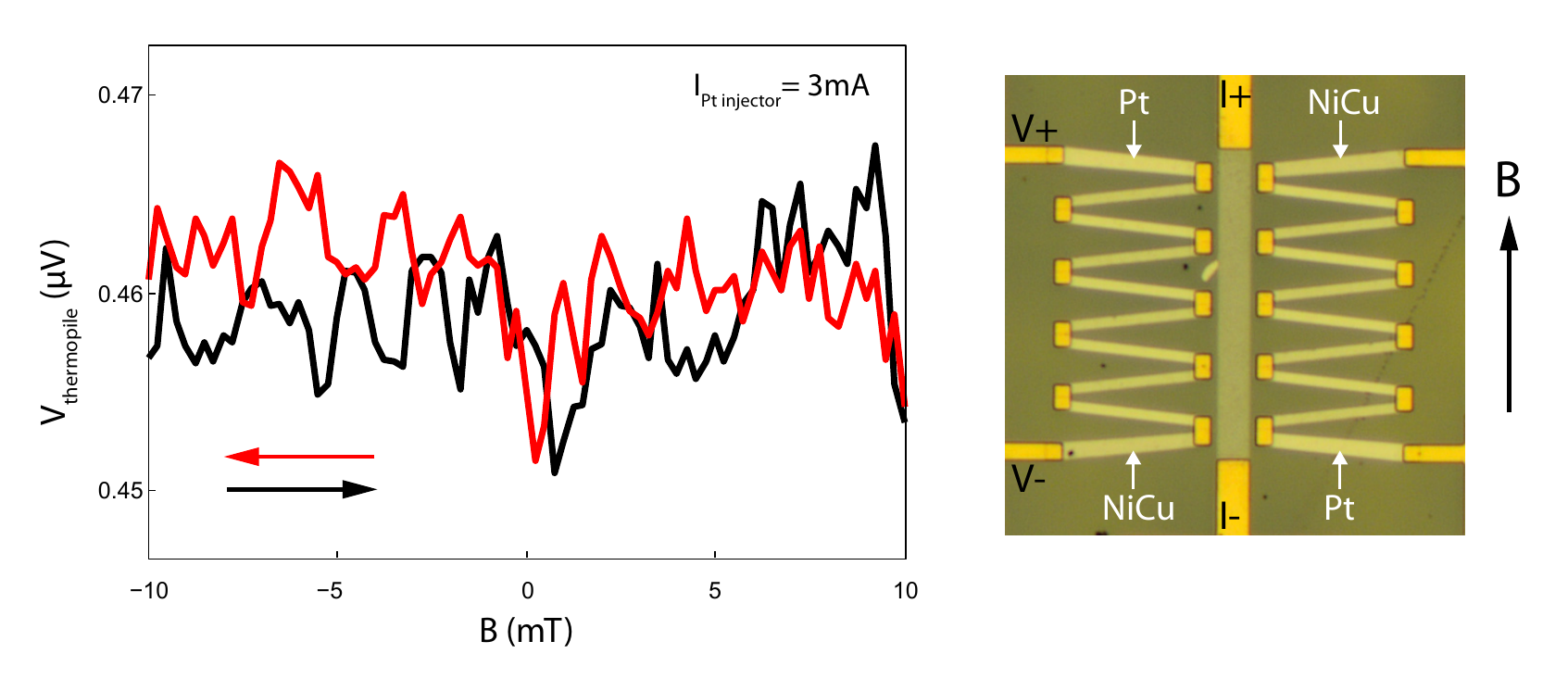}
\caption{(a) First harmonic voltage across the thermopile as a function of
applied magnetic field parallel to the Pt injector. (b) Optical microscope
picture of the measured device and measurement geometry. }
\label{fig:B_parallel}
\end{figure}

We repeated the measurements in the main text after rotating the magnetic
field by 90$^\circ$ (see Fig. \ref{fig:B_parallel}). The magnetization
direction is here parallel to the current in the Pt film such that the
current-induced spin accumulation is normal to the magnetization. The
background voltage and noise level remain unmodified; we do not detect a
heating or cooling of the ferromagnet. This confirms our interpretation of
the experiments in the main text. The spin torque normal to the
magnetization is not expected to affect the magnon temperature and the SPE
should vanish, as observed.

\section{III. T\lowercase{he 3}D \lowercase{finite element model}}

We adopt the three reservoir model of thermalized electron, magnon, and phonon systems. Charges are transported by the electron system only, while heat currents flow in all subsystems. We take into account spin angular momentum currents in the electron and magnon system, but disregard the phonon angular momentum current. The bulk charge and
heat currents in linear response are given by Eq. (4) in the main text. The charge and energy conservation relations read: 
\begin{equation}
\renewcommand\arraystretch{2.1}
\vec{\nabla} \cdot \begin{pmatrix} \vec{J}_{c} \\ \vec{Q}_{ph} \\ \vec{Q}_{m} \\ \vec{Q}_{e} \end{pmatrix}= \begin{pmatrix} 0 \\ \frac{\left(1-P_{m\mhyphen ph}^2\right)\left(\kappa_{m}+\kappa_{ph}\right)}{4\lambda_{m\mhyphen ph}^2} (T_{m}-T_{ph}) + \frac{\left(1-P_{e\mhyphen ph}^2\right)\left(\kappa_{e}+\kappa_{ph}\right)}{4\lambda_{e\mhyphen ph}^2} (T_{e}-T_{ph}) \\ 
-\frac{\left(1-P_{m\mhyphen ph}^2\right)\left(\kappa_{m}+\kappa_{ph}\right)}{4\lambda_{m\mhyphen ph}^2} (T_{m}-T_{ph}) \\ 
\frac{J_{c}^2}{\sigma}-\frac{\left(1-P_{e\mhyphen ph}^2\right)\left(\kappa_{e}+\kappa_{ph}\right)}{4\lambda_{e\mhyphen ph}^2} (T_{e}-T_{ph}) \end{pmatrix}
\label{eq:model_source}
\end{equation} with $P_{m-ph}=\left( \kappa _{m}-\kappa _{ph}\right) /\left( \kappa
_{m}+\kappa _{ph}\right) $, $P_{e-ph}=\left( \kappa _{e}-\kappa _{ph}\right)
/\left( \kappa _{e}+\kappa _{ph}\right) \ $and $J_{c}^{2}/\sigma $\vspace{%
0.1cm} accounts for Joule heating. The other terms describe the heat
exchange between phonons, magnons and electrons, as indicated by the
subscripts. All parameters have been defined in the main text, except for
the thermal relaxation lengths $\lambda _{m-ph}$ and $\lambda _{e-ph}.$ The
former is used as an adjustable parameter while the latter $\lambda _{e-ph}=%
\sqrt{\kappa _{e}/g}$ is calculated from the electron-phonon coupling
parameter ($g$) for Au and Pt \cite{EPrel}. This leads to a $\lambda
_{e-ph}^{Pt}$ of 4.5 nm and a $\lambda _{e-ph}^{Au}$ of 80 nm at room
temperature.

The boundary conditions at the metal$|$YIG interfaces are governed by the
phonon ($\kappa _{ph}^{I}$) and magnetic ($\varkappa _{S}^{I}$) interface
heat conductances. We adopt the values for $\kappa _{ph}^{I}$ from Ref. %
\onlinecite{SchreierSSE}, while that for Ni$_{45}$Cu$_{55}$$|$YIG is assumed
to be equal to that of Pt$|$YIG. The interface magnetic heat conductance is
calculated using \cite{SchreierSSE} 
\begin{equation}
\varkappa _{S}^{I}=\dfrac{h}{e^{2}}\dfrac{k_{B}T}{\hbar }\dfrac{\mu
_{B}k_{B}g_{r}}{\pi M_{s}V_{2m}}\left( 1+g_{S}\rho \lambda \coth \left( 
\dfrac{t}{\lambda }\right) \right) ^{-1}
\end{equation}%
where $V_{2m}$ is the (spin pumping) magnetic coherence volume. The currents
through the interface read 
\begin{equation}
\renewcommand\arraystretch{2}
\begin{pmatrix} J^{I}_{c} \\ Q^{I}_{ph} \\ Q^{I}_{m} \\ Q^{I}_{e} \end{pmatrix}= 
\begin{pmatrix} 0 \\ 
\kappa^I_{ph}(T_{ph}^{a}-T_{ph}^{b}) \\ 
-\varkappa _{S}^{I}(T_{m}-T_{e}) -  L_{S}T\dfrac{V_{s}}{2}\\ 
\varkappa _{S}^{I}(T_{m}-T_{e}) +  L_{S}T\dfrac{V_{s}}{2} \end{pmatrix}
\label{eq:model_fluxSPE}
\end{equation} where $T_{ph}^{\textrm{Pt/YIG}}$ are the phonon temperatures on the Pt/YIG side of
the interface. $Q_{m}^{I}$ and $Q_{e}^{I}$ are the interface magnetic heat currents. 
The first terms on the r.h.s represent the heat current driven by temperature differences,
while the second term is the heat current associated with the magnon
injection by an applied $V_{s}$ (see Eq. (2) in the main text) that is
responsible for the SPE.

Eq. (4) in the main text is solved for the SPE
and SSE configurations, taking into account energy conservation (Eq. (\ref%
{eq:model_source})) and boundary conditions (Eq. (\ref{eq:model_fluxSPE})).
The results are plotted in Figs. 3 (a) and (b) of the main text.

\section{IV. S\lowercase{pin} S\lowercase{eebeck effect}}

\begin{figure}[t]
\includegraphics{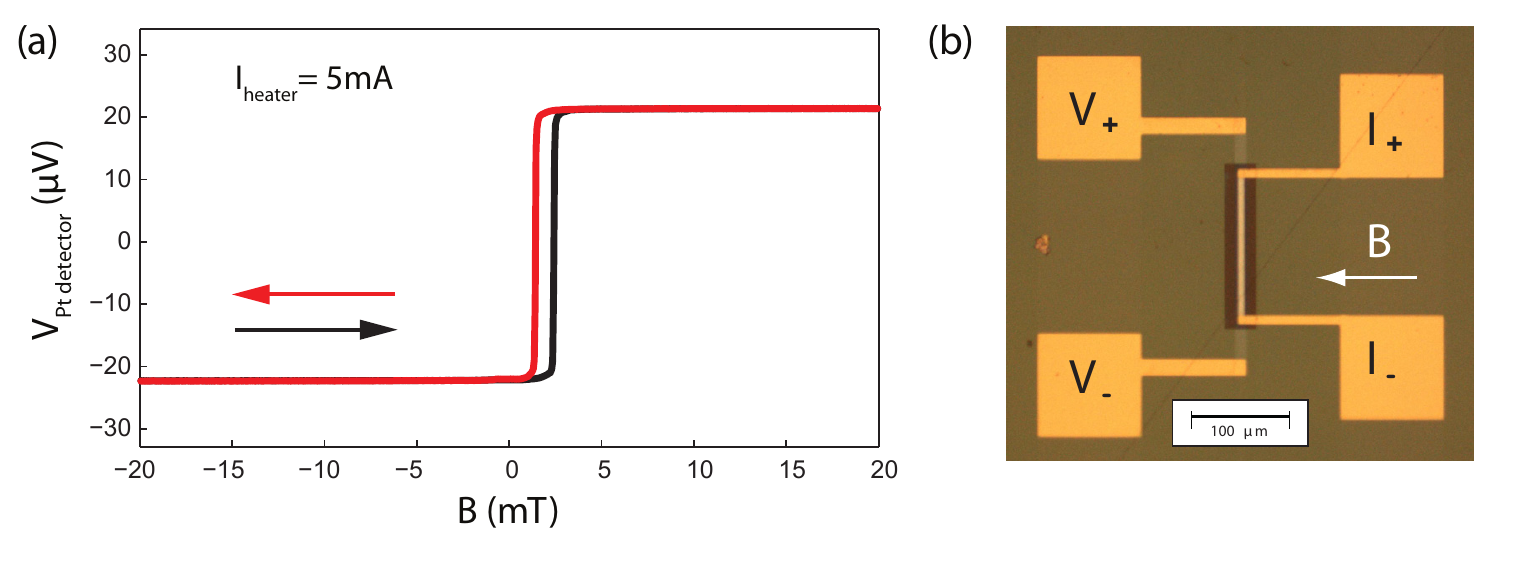}
\caption{(a) Second harmonic voltage across the Pt detector as a function of
applied magnetic field. (b) Optical microscope picture of the measured
device and the measurement geometry used. }
\label{fig:SSE}
\end{figure}

\begin{figure}[b]
\includegraphics{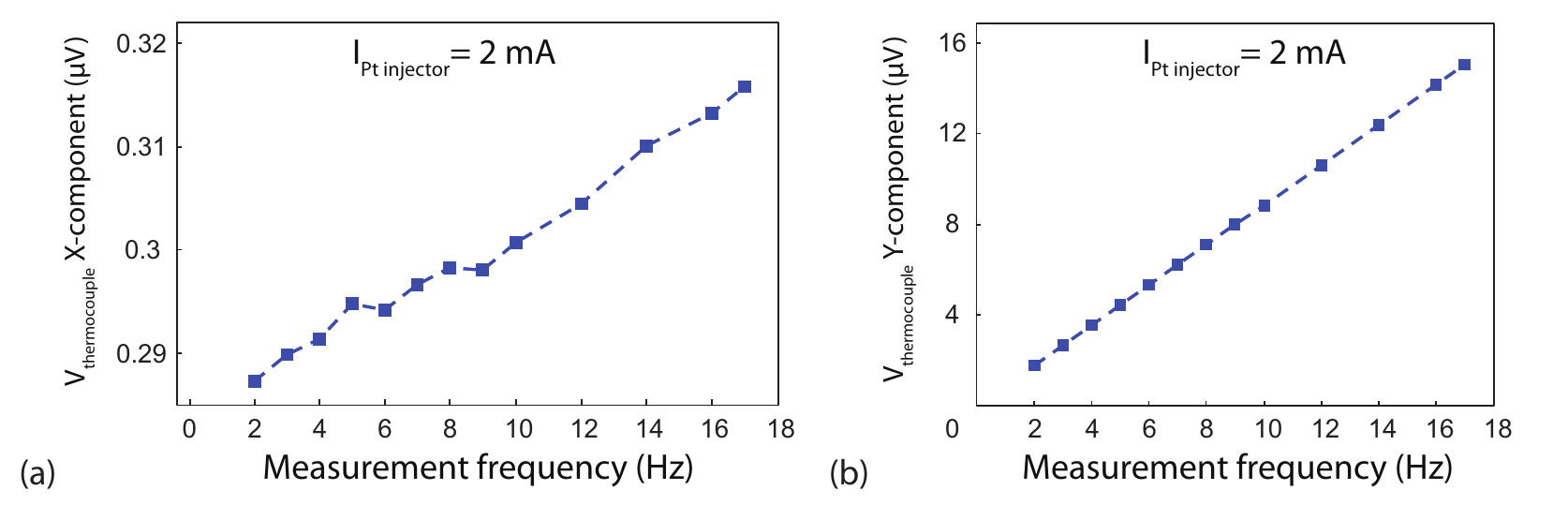}
\caption{(a) First harmonic voltage across the thermopile in phase with the
current ($X$-component). (b) First harmonic voltage across the thermopile
out-of-phase with the current ($Y$-component). }
\label{fig:Fdependence}
\end{figure}

To verify reciprocity we fabricated samples nominally
identical to the Pt$|$YIG heterostructures used for the SPE experiments in
order to measure the longitudinal SSE \cite{UchidaSSEl,WeilerSSEl}. Fig. %
\ref{fig:SSE}(b) gives a picture of such a device consisting of a 5 nm thick
sputtered Pt detector (250 x 10 $\mathrm{\mu }$m$^{2}$) on top of a GGG$|$%
YIG substrate as used for the study of the SPE effect. The Pt detector is
covered with a $\sim $70 nm aluminium oxide (Al$_{2}$O$_{3}$) layer that
electrically isolates the Pt detector from a 40 nm Pt film heater evaporated
on top. Both the detector and heater are contacted by a 100 nm thick Au
layer to large bonding pads.

By Joule heating, a charge current through the heater creates a thermal
gradient over the Pt$|$YIG interface. The hot electrons transfer energy to
the cold magnons, which is associated with a spin current in Pt that is
converted to an observable charge current by the inverse spin Hall effect.
This is the SSE. 

In Fig. \ref{fig:SSE}(a) the second harmonic voltage versus magnetic field
is a measure of the Joule heating. A clear SSE signal is observed, which
changes sign when the magnetization is reversed, as expected. The signal
scales quadratically with the current and for B parallel to the Pt detector
no SSE signal is detected, which confirms that the voltage is due to the
inverse spin Hall effect.

The model discussed in the main text and the previous section can be applied
to this measurement geometry to find the $T_{m-e}$ at the interface and the transverse
voltage over the Pt detector \cite{SchreierSSE}: 
\begin{equation}
V_{SSE}=2\cdot \frac{g_{r}\gamma \hbar k_{B}}{2\pi M_{s}V_{2m}}T_{m-e}\cdot 
\frac{4\pi }{e}\theta \rho l\cdot \dfrac{\frac{\lambda }{t}\tanh \left( 
\frac{t}{2\lambda }\right) }{1+g_{S}\rho \lambda \coth \left( t/\lambda
\right) }  \label{SSE_voltage}
\end{equation}%
where $l$ is the length of the Pt detector. The results obtained from the
SSE modeling are shown in Fig. 3(b) of the main text.

\section{V. F\lowercase{requency dependent measurements}}
We measured the voltage as function of frequency in order to exclude any
capacitive or inductive coupling between the Pt injector and the thermopile
(see Fig. \ref{fig:Fdependence}). The in-phase voltage ($x$-component)
slightly decreases with  frequency, leading to a frequency-dependent voltage
of around 10\% at our usual measurement frequency of 17 Hz (Fig. \ref%
{fig:Fdependence}(a)). The out-of-phase voltage ($Y$-component) linearly
depends on frequency and vanishes for zero frequency, as expected.

\begin{figure}[t]
\includegraphics{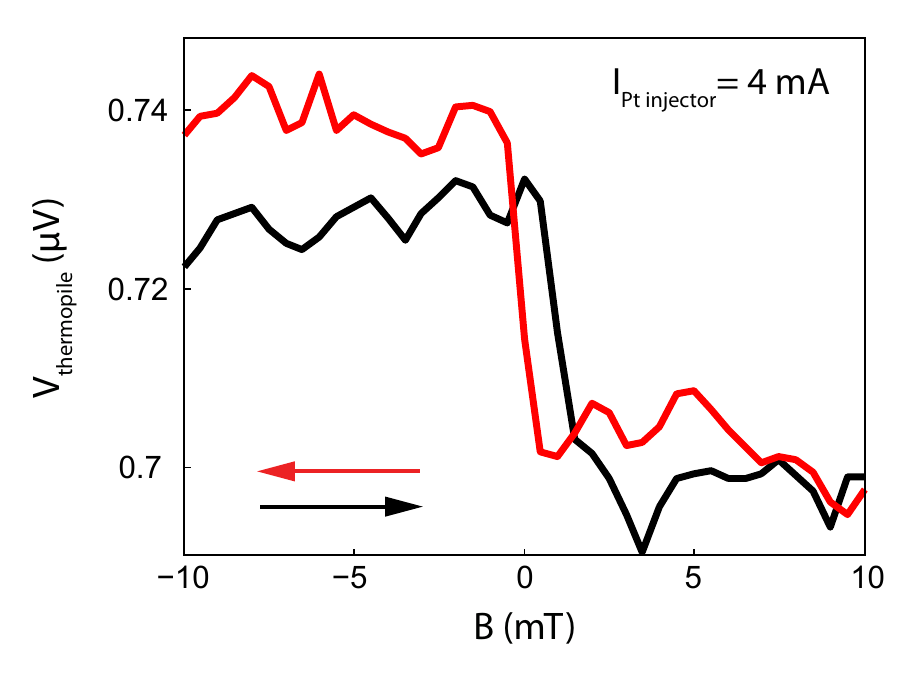}
\caption{First harmonic voltage across the thermopile as a function of
applied magnetic field for a 4 mA current through the Pt injector with
frequency of 3 Hz. }
\label{fig:3Hz}
\end{figure}

In Fig. \ref{fig:3Hz} the first harmonic voltage across the thermopile is
shown for a 4 mA current through the Pt injector with a frequency of 3 Hz. A
clear SPE signal is observed with the same magnitude as for the measurements
at 17 Hz . From these results we can safely conclude that the measured
signal is not affected by any spurious capacitive or inductive signals. Furthermore at these low frequencies the thermal time constants are much shorter than the period of the measurement modulation.

\begin{thebibliography}{20}%
\makeatletter
\providecommand \@ifxundefined [1]{%
 \@ifx{#1\undefined}
}%
\providecommand \@ifnum [1]{%
 \ifnum #1\expandafter \@firstoftwo
 \else \expandafter \@secondoftwo
 \fi
}%
\providecommand \@ifx [1]{%
 \ifx #1\expandafter \@firstoftwo
 \else \expandafter \@secondoftwo
 \fi
}%
\providecommand \natexlab [1]{#1}%
\providecommand \enquote  [1]{``#1''}%
\providecommand \bibnamefont  [1]{#1}%
\providecommand \bibfnamefont [1]{#1}%
\providecommand \citenamefont [1]{#1}%
\providecommand \href@noop [0]{\@secondoftwo}%
\providecommand \href [0]{\begingroup \@sanitize@url \@href}%
\providecommand \@href[1]{\@@startlink{#1}\@@href}%
\providecommand \@@href[1]{\endgroup#1\@@endlink}%
\providecommand \@sanitize@url [0]{\catcode `\\12\catcode `\$12\catcode
  `\&12\catcode `\#12\catcode `\^12\catcode `\_12\catcode `\%12\relax}%
\providecommand \@@startlink[1]{}%
\providecommand \@@endlink[0]{}%
\providecommand \url  [0]{\begingroup\@sanitize@url \@url }%
\providecommand \@url [1]{\endgroup\@href {#1}{\urlprefix }}%
\providecommand \urlprefix  [0]{URL }%
\providecommand \Eprint [0]{\href }%
\providecommand \doibase [0]{http://dx.doi.org/}%
\providecommand \selectlanguage [0]{\@gobble}%
\providecommand \bibinfo  [0]{\@secondoftwo}%
\providecommand \bibfield  [0]{\@secondoftwo}%
\providecommand \translation [1]{[#1]}%
\providecommand \BibitemOpen [0]{}%
\providecommand \bibitemStop [0]{}%
\providecommand \bibitemNoStop [0]{.\EOS\space}%
\providecommand \EOS [0]{\spacefactor3000\relax}%
\providecommand \BibitemShut  [1]{\csname bibitem#1\endcsname}%
\let\auto@bib@innerbib\@empty
\bibitem [{\citenamefont {Uchida}\ \emph {et~al.}(2010)\citenamefont {Uchida},
  \citenamefont {Xiao}, \citenamefont {Adachi}, \citenamefont {Ohe},
  \citenamefont {Takahashi}, \citenamefont {Ieda}, \citenamefont {Ota}, \citenamefont {Kajiwara}, \citenamefont {Umezawa},  	    \citenamefont {Kawai}, \citenamefont {Bauer}, \citenamefont {Maekawa} \
  and\ \citenamefont {Saitoh}}]{UchidaInsulator}%
  \BibitemOpen
  \bibfield  {author} {\bibinfo {author} {\bibfnamefont {K.}~\bibnamefont
  {Uchida}}, \bibinfo {author} {\bibfnamefont {J.}~\bibnamefont {Xiao}},
  \bibinfo {author} {\bibfnamefont {H.}~\bibnamefont {Adachi}}, \bibinfo
  {author} {\bibfnamefont {J. I.}~\bibnamefont {Ohe}}, \bibinfo {author}
  {\bibfnamefont {S.}~\bibnamefont {Takahashi}}, \bibinfo {author}
  {\bibfnamefont {J.}~\bibnamefont {Ieda}},  \bibinfo {author}
  {\bibfnamefont {T.}~\bibnamefont {Ota}},  \bibinfo {author}
  {\bibfnamefont {Y.}~\bibnamefont {Kajiwara}},  \bibinfo {author}
  {\bibfnamefont {H.}~\bibnamefont {Umezawa}},  \bibinfo {author}
  {\bibfnamefont {H.}~\bibnamefont {Kawai}},  \bibinfo {author}
  {\bibfnamefont {G. E. W.}~\bibnamefont {Bauer}},  \bibinfo {author}
  {\bibfnamefont {S.}~\bibnamefont {Maekawa}}, \ and\ \bibinfo {author}
  {\bibfnamefont {E.}~\bibnamefont {Saitoh}},\ }\href {\doibase 10.1038/nmat2856}
  {\bibfield  {journal} {\bibinfo  {journal} {Nature Materials}\ }\textbf {\bibinfo
  {volume} {9}},\ \bibinfo {pages} {894} (\bibinfo {year}
  {2010})}\BibitemShut {NoStop}%
\bibitem [{\citenamefont {Xiao}\ \emph {et~al.}(2010)
  \citenamefont {Xiao},
  \citenamefont {Bauer}, 
  \citenamefont {Uchida}, 
  \citenamefont {Saitoh},
  \citenamefont {Maekawa} \
  and\ \citenamefont {Xiao}}]{TheoryXiaoSSE}%
  \BibitemOpen
  \bibfield  {author} {
  \bibinfo {author} {\bibfnamefont {J.}~\bibnamefont {Xiao}}, 
  \bibinfo {author} {\bibfnamefont {G. E. W.}~\bibnamefont {Bauer}}, 
  \bibinfo {author} {\bibfnamefont {K.}~\bibnamefont {Uchida}}, 
  \bibinfo {author} {\bibfnamefont {E.}~\bibnamefont {Saitoh}}, 
  \bibinfo {author} {\bibfnamefont {S.}~\bibnamefont {Maekawa}}, \ }\href {\doibase 10.1103/PhysRevB.81.214418}
  {\bibfield  {journal} {\bibinfo  {journal} {Phys. Rev. B}\ }\textbf {\bibinfo
  {volume} {81}},\ \bibinfo {pages} {214418} (\bibinfo {year}
  {2010})}\BibitemShut {NoStop}%
\bibitem [{\citenamefont {Adachi}\ \emph {et~al.}(2011)
  \citenamefont {Adachi},
  \citenamefont {Ohe}, 
  \citenamefont {Takahashi}, 
  \citenamefont {Maekawa} \
  and\ \citenamefont {Xiao}}]{TheoryAdachiSSE}%
  \BibitemOpen
  \bibfield  {author} {
  \bibinfo {author} {\bibfnamefont {H.}~\bibnamefont {Adachi}}, 
  \bibinfo {author} {\bibfnamefont {J. I.}~\bibnamefont {Ohe}}, 
  \bibinfo {author} {\bibfnamefont {S.}~\bibnamefont {Takahashi}}, 
  \bibinfo {author} {\bibfnamefont {S.}~\bibnamefont {Maekawa}}, \ }\href {\doibase 10.1103/PhysRevB.83.094410}
  {\bibfield  {journal} {\bibinfo  {journal} {Phys. Rev. B}\ }\textbf {\bibinfo
  {volume} {83}},\ \bibinfo {pages} {094410} (\bibinfo {year}
  {2011})}\BibitemShut {NoStop}%
\bibitem [{\citenamefont {Hoffman}\ \emph {et~al.}(2013)
  \citenamefont {Hoffman},
  \citenamefont {Sato}, 
  \citenamefont {Tserkovnyak} \
  and\ \citenamefont {Hoffman}}]{TheoryTserkSSE}%
  \BibitemOpen
  \bibfield  {author} {
  \bibinfo {author} {\bibfnamefont {S.}~\bibnamefont {Hoffman}}, 
  \bibinfo {author} {\bibfnamefont {K.}~\bibnamefont {Sato}}, 
  \bibinfo {author} {\bibfnamefont {Y.}~\bibnamefont {Tserkovnyak}}, \ }\href {\doibase 10.1103/PhysRevB.88.064408}
  {\bibfield  {journal} {\bibinfo  {journal} {Phys. Rev. B}\ }\textbf {\bibinfo
  {volume} {88}},\ \bibinfo {pages} {064408} (\bibinfo {year}
  {2013})}\BibitemShut {NoStop}%
\bibitem [{\citenamefont {Hoffmann}\ \emph {et~al.}(2013)
  \citenamefont {Hoffmann} \
  and\ \citenamefont {Hoffmann}}]{SHE}%
  \BibitemOpen
  \bibfield  {author} {
  \bibinfo {author} {\bibfnamefont {A.}~\bibnamefont {Hoffmann}}, \ }\href {\doibase : 10.1109/TMAG.2013.2262947}
  {\bibfield  {journal} {\bibinfo  {journal} {IEEE Trans. Magnetics}\ }\textbf {\bibinfo
  {volume} {49}},\ \bibinfo {pages} {5172} (\bibinfo {year}
  {2013})}\BibitemShut {NoStop}%
\bibitem [{\citenamefont {Schreier}\ \emph {et~al.}(2013)
  \citenamefont {Schreier},
  \citenamefont {Kamra}, 
  \citenamefont {Weiler}, 
  \citenamefont {Xiao},
  \citenamefont {Bauer},
  \citenamefont {Gross},
  \citenamefont {Goennenwein} \
  and\ \citenamefont {Schreier}}]{SchreierSSE}%
  \BibitemOpen
  \bibfield  {author} {
  \bibinfo {author} {\bibfnamefont {M.}~\bibnamefont {Schreier}}, 
  \bibinfo {author} {\bibfnamefont {A.}~\bibnamefont {Kamra}}, 
  \bibinfo {author} {\bibfnamefont {M.}~\bibnamefont {Weiler}}, 
  \bibinfo {author} {\bibfnamefont {J.}~\bibnamefont {Xiao}},
  \bibinfo {author} {\bibfnamefont {G. E. W.}~\bibnamefont {Bauer}},
  \bibinfo {author} {\bibfnamefont {R.}~\bibnamefont {Gross}}, 
  \bibinfo {author} {\bibfnamefont {S. T. B.}~\bibnamefont {Goennenwein}},\ } \href {\doibase 10.1103/PhysRevB.88.094410}
  {\bibfield  {journal} {\bibinfo  {journal} {Phys. Rev. B}\ }\textbf {\bibinfo
  {volume} {88}},\ \bibinfo {pages} {094410} (\bibinfo {year}
  {2013})}\BibitemShut {NoStop}%
\bibitem [{\citenamefont {Onsager}\ \emph {et~al.}(1931)
  \citenamefont {Onsager}, \
  and\ \citenamefont {Onsager}}]{Onsager}%
  \BibitemOpen
  \bibfield  {author} {
  \bibinfo {author} {\bibfnamefont {L.}~\bibnamefont {Onsager}},  \ }\href {\doibase 10.1103/PhysRev.37.405}
  {\bibfield  {journal} {\bibinfo  {journal} {Phys. Rev.}\ }\textbf {\bibinfo
  {volume} {37}},\ \bibinfo {pages} {405–426} (\bibinfo {year}
  {1931})}\BibitemShut {NoStop}%
\bibitem [{\citenamefont {Agrawal}\ \emph {et~al.}(2013)
  \citenamefont {Agrawal},
  \citenamefont {Vasyuchka}, 
  \citenamefont {Serga}, 
  \citenamefont {Karenowska},
  \citenamefont {Melkov},
  \citenamefont {Hillebrands} \
  and\ \citenamefont {Agrawal}}]{TmpHillebrands}%
  \BibitemOpen
  \bibfield  {author} {
  \bibinfo {author} {\bibfnamefont {M.}~\bibnamefont {Agrawal}}, 
  \bibinfo {author} {\bibfnamefont {V. I.}~\bibnamefont {Vasyuchka}}, 
  \bibinfo {author} {\bibfnamefont {A. A.}~\bibnamefont {Serga}}, 
  \bibinfo {author} {\bibfnamefont {A. D.}~\bibnamefont {Karenowska}},
  \bibinfo {author} {\bibfnamefont {G. A.}~\bibnamefont {Melkov}},
  \bibinfo {author} {\bibfnamefont {B.}~\bibnamefont {Hillebrands}}, \ } \href {\doibase 10.1103/PhysRevLett.111.107204}
  {\bibfield  {journal} {\bibinfo  {journal} {Phys. Rev. Lett.}\ }\textbf {\bibinfo
  {volume} {111}},\ \bibinfo {pages} {107204} (\bibinfo {year}
  {2013})}\BibitemShut {NoStop}%
\bibitem [{\citenamefont {Roschewsky}\ \emph {et~al.}(2013)
  \citenamefont {Roschewsky}, \
  and\ \citenamefont {Roschewsky}}]{TmpRoschewsky}%
  \BibitemOpen
  \bibfield  {author} {
  \bibinfo {author} {\bibfnamefont {N.}~\bibnamefont {Roschewsky}}, 
  \bibinfo {author} {\bibfnamefont {M.}~\bibnamefont {Schreier}}, 
  \bibinfo {author} {\bibfnamefont {A.}~\bibnamefont {Kamra}}, 
  \bibinfo {author} {\bibfnamefont {F.}~\bibnamefont {Schade}},
  \bibinfo {author} {\bibfnamefont {K.}~\bibnamefont {Ganzhorn}},
  \bibinfo {author} {\bibfnamefont {S.}~\bibnamefont {Meyer}},
  \bibinfo {author} {\bibfnamefont {H.}~\bibnamefont {Huebl}},
  \bibinfo {author} {\bibfnamefont {S.}~\bibnamefont {Gepr\"{a}gs}},
  \bibinfo {author} {\bibfnamefont {R.}~\bibnamefont {Gross}},
  \bibinfo {author} {\bibfnamefont {S. T. B.}~\bibnamefont {Goennenwein}}, \ } \href {http://arxiv.org/abs/1309.3986v1}
  {\bibfield  {journal} {\bibinfo  {journal} {arXiv 1309.3986v1}\ } (\bibinfo {year}
  {2013})}\BibitemShut {NoStop}%
\bibitem [{\citenamefont {Brataas}\ \emph {et~al.}(2006)
  \citenamefont {Brataas},
  \citenamefont {Bauer},
  \citenamefont {Kelly} \
  and\ \citenamefont {Brataas}}]{MixC}%
  \BibitemOpen
  \bibfield  {author} {
  \bibinfo {author} {\bibfnamefont {A.}~\bibnamefont {Brataas}}, 
  \bibinfo {author} {\bibfnamefont {G. E. W.}~\bibnamefont {Bauer}},
  \bibinfo {author} {\bibfnamefont {P. J.}~\bibnamefont {Kelly}}, \ } \href {http://www.sciencedirect.com/science/article/pii/S0370157306000238}
  {\bibfield  {journal} {\bibinfo  {journal} {Phys. Rep.}\ }\textbf {\bibinfo
  {volume} {427}},\ \bibinfo {pages} {157} (\bibinfo {year}
  {2006})}\BibitemShut {NoStop}%
\bibitem [{\citenamefont {Wang}\ (2011)
  \citenamefont {Wang} \
  and\ \citenamefont {Wang}}]{Wang}%
  \BibitemOpen
  \bibfield  {author} {
  \bibinfo {author} {\bibfnamefont {Z.}~\bibnamefont {Wang}},
  \bibinfo {author} {\bibfnamefont {Y.}~\bibnamefont {Sun}},
  \bibinfo {author} {\bibfnamefont {M.}~\bibnamefont {Wu}}, 
  \bibinfo {author} {\bibfnamefont {V.}~\bibnamefont {Tiberkevich}},
  \bibinfo {author} {\bibfnamefont {A.}~\bibnamefont {Slavin}},\ }\href {\doibase 10.1103/PhysRevLett.107.146602}
  {\bibfield  {journal} {\bibinfo  {journal} {Phys. Rev. Lett.}\ }\textbf {\bibinfo
  {volume} {107}},\ \bibinfo {pages} {146602} (\bibinfo {year}
  {2011})}\BibitemShut {NoStop}%
\bibitem [{\citenamefont {Callen}\ (1948)
  \citenamefont {Callen} \
  and\ \citenamefont {Callen}}]{OnsagerR}%
  \BibitemOpen
  \bibfield  {author} {
  \bibinfo {author} {\bibfnamefont {H. B.}~\bibnamefont {Callen}}, \ }\href {\doibase 10.1103/PhysRev.73.1349}
  {\bibfield  {journal} {\bibinfo  {journal} {Phys. Rev.}\ }\textbf {\bibinfo
  {volume} {73}},\ \bibinfo {pages} {1349} (\bibinfo {year}
  {1948})}\BibitemShut {NoStop}%
\bibitem {G_S}%
  \BibitemOpen
  \bibfield  {author} {
  \bibinfo {author} {\bibfnamefont {H. J.}~\bibnamefont {Jiao}}, 
  \bibinfo {author} {\bibfnamefont {J.}~\bibnamefont {Xiao}},
  \bibinfo {author} {\bibfnamefont {G. E. W.}~\bibnamefont {Bauer}},\ }
  in preparation\BibitemShut {NoStop}%
\bibitem [{\citenamefont {Vlietstra}\ (2013)
  \citenamefont {Vlietstra} \
  and\ \citenamefont {Vlietstra}}]{VSM}%
  \BibitemOpen
  \bibfield  {author} { 
  \bibinfo {author} {\bibfnamefont {N.}~\bibnamefont {Vlietstra}},
  \bibinfo {author} {\bibfnamefont {J.}~\bibnamefont {Shan}},
  \bibinfo {author} {\bibfnamefont {V.}~\bibnamefont {Castel}},
  \bibinfo {author} {\bibfnamefont {B. J.}~\bibnamefont {van Wees}},
  \bibinfo {author} {\bibfnamefont {J.}~\bibnamefont {Ben Youssef}},\ } \href {\doibase 10.1103/PhysRevB.87.184421}
  {\bibfield  {journal} {\bibinfo  {journal} {Phys. Rev. B}\ }\textbf {\bibinfo
  {volume} {87}},\ \bibinfo {pages} {184421} (\bibinfo {year}
  {2013})}\BibitemShut {NoStop}%
\bibitem {comsol}%
  \BibitemOpen
  COMSOL Multiphysics$^{\textrm{\textregistered}}$ 4.2a \BibitemShut {NoStop}%
\bibitem [{\citenamefont {Vlietstra}\ (2013)
  \citenamefont {Vlietstra} \
  and\ \citenamefont {Vlietstra}}]{MixCMeas}%
  \BibitemOpen
  \bibfield  {author} { 
  \bibinfo {author} {\bibfnamefont {N.}~\bibnamefont {Vlietstra}},
  \bibinfo {author} {\bibfnamefont {J.}~\bibnamefont {Shan}},
  \bibinfo {author} {\bibfnamefont {V.}~\bibnamefont {Castel}},
  \bibinfo {author} {\bibfnamefont {J.}~\bibnamefont {Ben Youssef}},
  \bibinfo {author} {\bibfnamefont {G. E. W.}~\bibnamefont {Bauer}},
  \bibinfo {author} {\bibfnamefont {B. J.}~\bibnamefont {van Wees}},\ } \href {\doibase 10.1063/1.4813760}
  {\bibfield  {journal} {\bibinfo  {journal} {Appl. Phys. Lett.}\ }\textbf {\bibinfo
  {volume} {103}},\ \bibinfo {pages} {032401} (\bibinfo {year}
  {2013})}\BibitemShut {NoStop}%
\bibitem [{\citenamefont {Douglass}\ (1963)
  \citenamefont {Douglass} \
  and\ \citenamefont {Douglass}}]{KMagnon}%
  \BibitemOpen
  \bibfield  {author} { 
  \bibinfo {author} {\bibfnamefont {R. L.}~\bibnamefont {Douglass}},\ } \href {\doibase 10.1103/PhysRev.129.1132}
  {\bibfield  {journal} {\bibinfo  {journal} {Phys. Rev.}\ }\textbf {\bibinfo
  {volume} {129}},\ \bibinfo {pages} {1132} (\bibinfo {year}
  {1963})}\BibitemShut {NoStop}%
\bibitem [{\citenamefont {Swartz}\ (1989)
  \citenamefont {Swartz} \
  and\ \citenamefont {Swartz}}]{IRes}%
  \BibitemOpen
  \bibfield  {author} { 
  \bibinfo {author} {\bibfnamefont {E. T.}~\bibnamefont {Swartz}},
  \bibinfo {author} {\bibfnamefont {R. O.}~\bibnamefont {Pohl}}, \ } \href {\doibase 10.1103/RevModPhys.61.605}
  {\bibfield  {journal} {\bibinfo  {journal} {Rev. Mod. Phys.}\ }\textbf {\bibinfo
  {volume} {61}},\ \bibinfo {pages} {605} (\bibinfo {year}
  {1989})}\BibitemShut {NoStop}%
\bibitem [{\citenamefont {Bakker}\ (2011)
  \citenamefont {Bakker} \
  and\ \citenamefont {Bakker}}]{SandSigma}%
  \BibitemOpen
  \bibfield  {author} { 
  \bibinfo {author} {\bibfnamefont {F. L.}~\bibnamefont {Bakker}},
  \bibinfo {author} {\bibfnamefont {J.}~\bibnamefont {Flipse}},
  \bibinfo {author} {\bibfnamefont {B. J.}~\bibnamefont {van Wees}},\ } \href {\doibase 10.1063/1.3703675}
  {\bibfield  {journal} {\bibinfo  {journal} {J. Appl. Phys.}\ }\textbf {\bibinfo
  {volume} {111}},\ \bibinfo {pages} {084306} (\bibinfo {year}
  {2012})}\BibitemShut {NoStop}%
\bibitem [{\citenamefont {Sandweg}\ \emph {et~al.} (2011)
  \citenamefont {Sandweg} \
  and\ \citenamefont {Sandweg}}]{SSE_SP_Hill}%
  \BibitemOpen
  \bibfield  {author} { 
  \bibinfo {author} {\bibfnamefont {C. W.}~\bibnamefont {Sandweg}},
  \bibinfo {author} {\bibfnamefont {Y.}~\bibnamefont {Kajiwara}}, 
  \bibinfo {author} {\bibfnamefont {A. V.}~\bibnamefont {Chumak}}, 
  \bibinfo {author} {\bibfnamefont {A. A.}~\bibnamefont {Serga}},
  \bibinfo {author} {\bibfnamefont {V. I.}~\bibnamefont {Vasyuchka}},
  \bibinfo {author} {\bibfnamefont {M. B.}~\bibnamefont {Jungfleisch}}, 
  \bibinfo {author} {\bibfnamefont {E.}~\bibnamefont {Saitoh}},
  \bibinfo {author} {\bibfnamefont {B.}~\bibnamefont {Hillebrands}}, \ } \href {\doibase 10.1103/PhysRevLett.106.216601}
  {\bibfield  {journal} {\bibinfo  {journal} {Phys. Rev. Lett.}\ }\textbf {\bibinfo
  {volume} {106}},\ \bibinfo {pages} {216601} (\bibinfo {year}
  {2011})}\BibitemShut {NoStop}%
\bibitem [{\citenamefont {Weiler}\ \emph {et~al.} (2013)
  \citenamefont {Weiler} \
  and\ \citenamefont {Weiler}}]{SSE_L_YIG_2}%
  \BibitemOpen
  \bibfield  {author} { 
  \bibinfo {author} {\bibfnamefont {M.}~\bibnamefont {Weiler}},
  \bibinfo {author} {\bibfnamefont {M.}~\bibnamefont {Althammer}}, 
  \bibinfo {author} {\bibfnamefont {M.}~\bibnamefont {Schreier}}, 
  \bibinfo {author} {\bibfnamefont {J.}~\bibnamefont {Lotze}},
  \bibinfo {author} {\bibfnamefont {M.}~\bibnamefont {Pernpeintner}},
  \bibinfo {author} {\bibfnamefont {S.}~\bibnamefont {Meyer}}, 
  \bibinfo {author} {\bibfnamefont {H.}~\bibnamefont {Huebl}},
  \bibinfo {author} {\bibfnamefont {R.}~\bibnamefont {Gross}},
  \bibinfo {author} {\bibfnamefont {A.}~\bibnamefont {Kamra}},
  \bibinfo {author} {\bibfnamefont {J.}~\bibnamefont {Xiao}},
  \bibinfo {author} {\bibfnamefont {Y. T.}~\bibnamefont {Chen}},
  \bibinfo {author} {\bibfnamefont {H. J.}~\bibnamefont {Jiao}},
  \bibinfo {author} {\bibfnamefont {G. E. W.}~\bibnamefont {Bauer}},
  \bibinfo {author} {\bibfnamefont {S. T. B.}~\bibnamefont {Goennenwein}}, \ } \href {\doibase 10.1103/PhysRevLett.111.176601}
  {\bibfield  {journal} {\bibinfo  {journal} {Phys. Rev. Lett.}\ }\textbf {\bibinfo
  {volume} {111}},\ \bibinfo {pages} {176601} (\bibinfo {year}
  {2013})}\BibitemShut {NoStop}%
\bibitem [{\citenamefont {Wang}\ (2012)
  \citenamefont {Wang} \
  and\ \citenamefont {Wang}}]{EPrel}%
  \BibitemOpen
  \bibfield  {author} {
  \bibinfo {author} {\bibfnamefont {W.}~\bibnamefont {Wang}},
  \bibinfo {author} {\bibfnamefont {D. G.}~\bibnamefont {Cahill}}, \ }\href {\doibase 10.1103/PhysRevLett.109.175503}
  {\bibfield  {journal} {\bibinfo  {journal} {Phys. Rev. Lett.}\ }\textbf {\bibinfo
  {volume} {109}},\ \bibinfo {pages} {175503} (\bibinfo {year}
  {2012})}\BibitemShut {NoStop}%
\bibitem [{\citenamefont {Uchida}\ (2010)
  \citenamefont {Uchida} \
  and\ \citenamefont {Uchida}}]{UchidaSSEl}%
  \BibitemOpen
  \bibfield  {author} {
  \bibinfo {author} {\bibfnamefont {K.}~\bibnamefont {Uchida}},
  \bibinfo {author} {\bibfnamefont {H.}~\bibnamefont {Adachi}},
  \bibinfo {author} {\bibfnamefont {T.}~\bibnamefont {Ota}},
  \bibinfo {author} {\bibfnamefont {H.}~\bibnamefont {Nakayama}},
  \bibinfo {author} {\bibfnamefont {S.}~\bibnamefont {Maekawa}},
  \bibinfo {author} {\bibfnamefont {E.}~\bibnamefont {Saitoh}}, \ }\href {\doibase 10.1063/1.3507386}
  {\bibfield  {journal} {\bibinfo  {journal} {Appl. Phys. Lett.}\ }\textbf {\bibinfo
  {volume} {97}},\ \bibinfo {pages} {172505} (\bibinfo {year}
  {2010})}\BibitemShut {NoStop}%
\bibitem [{\citenamefont {Weiler}\ (2012)
  \citenamefont {Weiler} \
  and\ \citenamefont {Weiler}}]{WeilerSSEl}%
  \BibitemOpen
  \bibfield  {author} {
  \bibinfo {author} {\bibfnamefont {M.}~\bibnamefont {Weiler}},
  \bibinfo {author} {\bibfnamefont {M.}~\bibnamefont {Althammer}},
  \bibinfo {author} {\bibfnamefont {F. D.}~\bibnamefont {Czeschka}},
  \bibinfo {author} {\bibfnamefont {H.}~\bibnamefont {Huebl}},
  \bibinfo {author} {\bibfnamefont {M. S.}~\bibnamefont {Wagner}},
  \bibinfo {author} {\bibfnamefont {M.}~\bibnamefont {Opel}},
  \bibinfo {author} {\bibfnamefont {I.}~\bibnamefont {Imort}},
  \bibinfo {author} {\bibfnamefont {G.}~\bibnamefont {Reiss}},
  \bibinfo {author} {\bibfnamefont {A.}~\bibnamefont {Thomas}},
  \bibinfo {author} {\bibfnamefont {R.}~\bibnamefont {Gross}},
  \bibinfo {author} {\bibfnamefont {S. T. B.}~\bibnamefont {Goennenwein}}, \ }\href {\doibase 10.1103/PhysRevLett.108.106602}
  {\bibfield  {journal} {\bibinfo  {journal} {Phys. Rev. Lett.}\ }\textbf {\bibinfo
  {volume} {108}},\ \bibinfo {pages} {106602} (\bibinfo {year}
  {2012})}\BibitemShut {NoStop}%
\end{thebibliography}
\end{document}